\documentclass[12pt,a4paper,titlepage]{article}
\usepackage{epsfig}
\usepackage{amssymb}
\usepackage{amsmath}

\newcommand{\expn}{{\mathbb E}}
\newcommand{\tp}{^{\rm T}}

\newtheorem{thm}{Theorem}

\renewcommand\baselinestretch{2}
\begin{document}
\title{[ To appear in Statistica Sinica]\\Nonparametric Conditional Inference for Regression Coefficients with Application to Configural Polysampling}
\author{Yvonne H.S. Ho\\ {\small\it e-mail: yvonneho@imperial.ac.uk}\\[-2ex] {\small\it Department of Mathematics, Imperial College, London, U.K.}
\and Stephen M.S. Lee\thanks{Partially supported by a grant from the Research Grants Council
of the Hong Kong Special Administrative Region, China (Project No.\ HKU 7104/01P).}
\\ {\small\it e-mail:
smslee@hkusua.hku.hk}\\[-2ex] {\small\it Department of Statistics and Actuarial Science, The University of Hong Kong,}\\[-2ex] {\small\it
Pokfulam Road, Hong Kong}}
\date{}
\maketitle

\begin{center}
\bf{ABSTRACT}
\end{center}
We consider inference procedures, conditional on an observed
ancillary statistic, for regression coefficients under a linear
regression setup where the unknown error distribution is specified
nonparametrically. We establish conditional asymptotic normality
of the regression coefficient estimators under regularity
conditions, and formally justify the approach of plugging in
kernel-type density estimators in conditional inference
procedures. Simulation results show that the approach yields
accurate conditional coverage probabilities when used for
constructing confidence intervals. The plug-in approach can be
applied in conjunction with configural polysampling to derive
robust conditional estimators adaptive to a confrontation of
contrasting scenarios. We demonstrate this by investigating the
conditional mean squared error of location estimators under
various confrontations in a simulation study, which successfully
extends configural polysampling to a nonparametric context.
\thispagestyle{empty}
\newpage
\setcounter{page}{1}

\section{Introduction}
The classical conditionality principle (Fisher (1934, 1935); Cox
and Hinkley (1974)) demands that statistical inference be made
relevant to the data at hand by conditioning on ancillary
statistics. Arguments for this are best seen from examples in Cox
and Hinkley ((1974), Ch.2). Further discussion can be found in
Barndorff-Nielsen (1978) and in Lehmann (1981). Under regression
models, the ancillary statistic takes the form of studentized
residuals. Conditional inference about regression coefficients has
been discussed by Fraser (1979), Hinkley (1978), DiCiccio (1988),
DiCiccio, Field and Fraser (1990), and Severini (1996), among
others. When the error density is completely specified,
approximate conditional inference can be made by  Monte Carlo
simulation or by using numerical integration techniques. The
procedure nevertheless becomes computationally intensive if the
parameter has a high dimension, in which case large-sample
approximations such as those proposed by DiCiccio (1988) and
DiCiccio, Field and Fraser (1990) may be necessary. In a
nonparametric context where the error density is unspecified,
conditional inference has not received much attention despite its
clear practical relevance. Fraser (1976) and Severini (1994)
tackle the special case of location models. Both suggest plugging
in kernel density estimates but provide no theoretical
justification for the approach nor any formal suggestion on the
choice of bandwidth. The need for sophisticated Monte Carlo or
numerical integration techniques endures, and the computational
cost is even more expensive than that required by the parametric
case. Details of the computational procedures can be found in
Severini (1994) and Seifu, Severini and Tanner (1999). In the
present paper we prove asymptotic consistency, conditional on the
ancillary statistic, of plugging in the kernel density estimator,
and derive the orders of bandwidths sufficient for ensuring such
consistency. Our proof also suggests a normal approximation to the
plug-in approach which is computationally much more efficient for
high-dimensional regression estimators.

Consideration of conditionality has motivated different notions of
robustness for regression models: see Fraser (1979), Barnard
(1981, 1983), Hinkley (1983) and Severini (1992, 1996).
Morgenthaler and Tukey (1991) propose a configural polysampling
technique for robust conditional inference, which compromises
results obtained separately from a confrontation of contrasting
error distributions and provides a global perspective for
robustness. Our plug-in approach extends configural polysampling
to a nonparametric context, substantially broadens the scope of
confrontation, and enhances the global nature of the robustness
attributed to the resulting inference procedure.

Section~2.1 describes the problem setting. Section~2.2 reviews a
bootstrap approach to unconditional inference for regression
coefficients. The case of conditional inference is treated in
Section~2.3. Section~3 investigates the asymptotics underlying the
plug-in approach. Section~4 reviews configural polysampling and
extends it to nonparametric confrontations by the plug-in
approach. Empirical results are given in Section~5. Section~6
concludes our findings. All proofs are given in the Appendix.

\section{Inference for regression coefficients}
\subsection{Problem setting}
Consider a linear regression model $Y_i = x_i\tp{\beta} +
\tilde\epsilon_i$, for $i = 1, \ldots, n$, where $x_i = (x_{i1},
\ldots, x_{ip})\tp$ is the vector of covariates, $\beta =
(\beta_1, \ldots,\beta_p)\tp$ is the vector of unknown regression
coefficients, and the random errors
$\tilde\epsilon_1,\ldots,\tilde\epsilon_n$ are independent and
identically distributed with density $f$ symmetric about 0. Write
$Y=(Y_1, \ldots, Y_n)\tp$, $X=[x_1,\ldots,x_n]\tp$ and
$\tilde\epsilon=(\tilde\epsilon_1,\ldots,\tilde\epsilon_n)\tp$.
Introduction of a scale parameter leads to a regression-scale
model under which $f(u)=f_0(u/\sigma)/\sigma$ for an unknown scale
$\sigma>0$, and a density $f_0$ with unit scale. In this case we
have
$\tilde\epsilon=\sigma\epsilon=\sigma(\epsilon_1,\ldots,\epsilon_n)\tp$,
for independent $\epsilon_1,\ldots,\epsilon_n$ distributed with
density $f_0$, so that $Y=X\beta+\sigma\epsilon$. Throughout the
paper we treat $\beta$ as the parameter of interest and $f$, or
equivalently, $(\sigma,f_0)$, as the nuisance parameter of
possibly infinite dimension.

Let $\hat\beta=\hat\beta(Y)$ be a location and scale equivariant
estimator of $\beta$ and, under the regression-scale model,
$\hat\sigma=\hat\sigma(Y)$ be a location invariant and scale
equivariant estimator of $\sigma$, so that
$\hat\beta(Xc+dy)=c+d\hat\beta(y)$ and
$\hat\sigma(Xc+dy)=|d|\hat\sigma(y)$ for any $(d,c,y)\in{\Bbb
R}\times{\Bbb R}^p\times{\Bbb R}^n$. For example, $\hat\beta$ may
be the least squares estimator and $\hat\sigma^2$ the mean squared
residuals. Define, for $i=1,\ldots,n$, $\tilde A_i=Y_i - x_i\tp
\hat\beta$ and $A_i=\tilde A_i/\hat\sigma$. We can easily show
that $\tilde A=(\tilde A_1, \ldots, \tilde A_n)\tp$ and $A=(A_1,
\ldots, A_n)\tp$ provide ancillary statistics under the regression
model with known $f$ and the regression-scale model with known
$f_0$, respectively. When $f_0$, and hence $f$, is unspecified,
exact conditional inference is not possible as the conditional
likelihood of $\beta$ depends in general on $f_0$. Adopting J\o
rgensen's (1993) notion of I-sufficiency, we see that $A$ is
I-sufficient for $f_0$, so that any relevant information about
$f_0$ is contained in $A$. The same applies to $\tilde A$ and $f$.
Such ancillary-informed knowledge about $f$ and $f_0$ forms the
basis for nonparametric estimation of the conditional likelihood
and facilitates nonparametric conditional inference in an
approximate sense.

\subsection{Unconditional inference: a bootstrap approach}
Under the regression-scale model, the distribution $G_T$ of $T=(\hat\beta-\beta)/\hat\sigma$
does not depend on $(\beta,\sigma)$ and provides a basis for unconditional inference when $f_0$ is known.
The same applies to the distribution $G_U$ of $U=\hat\beta-\beta$
under the regression model.
Suppose now $f_0$, and hence $f$, is unspecified except for symmetry about 0. Under the regression-scale model,
we may estimate $G_T$ by the residual bootstrap method as follows.
Let $F_n$ be the empirical distribution of the $2n$ residuals $\pm A_1,\ldots,\pm A_n$.
For a random sample $\epsilon^*=(\epsilon^*_1,\ldots,\epsilon^*_n)\tp$ drawn from $F_n$,
construct a bootstrap resample $Y^{*} =X{\hat\beta} + \hat\sigma \epsilon^{*}$ and calculate $\hat\beta^*=\hat\beta(Y^*)$ and $\hat\sigma^*=\hat\sigma(Y^*)$.
The distribution $G_T$ is then estimated by the bootstrap distribution, $\hat{G}_T$ say, of $(\hat\beta^*-\hat\beta)/\hat\sigma^*$.
Under the regression model, we replace $A$ by $\tilde{A}$, calculate $\hat\beta^*$ from the
bootstrap resample $Y^{*} =X{\hat\beta} + \epsilon^{*}$ and estimate $G_U$ by the bootstrap
distribution $\hat G_{U}$ of $\hat\beta^*-\hat\beta$.

\subsection{Conditional inference: a plug-in approach}
Conditional inference about $\beta$ replaces $G_T$ and $G_U$ used
in the unconditional approach by, respectively, the conditional
distributions $G_{T|A}(\cdot|a)$ of $T$ given
$A=a=(a_1,\ldots,a_n)\tp$ and $G_{U|\tilde A}(\cdot|\tilde a)$ of
$U$ given $\tilde A=\tilde a=(\tilde a_1,\ldots,\tilde a_n)\tp$.

Consider first the regression-scale model. Define $S = \hat\sigma / \sigma$. The conditional joint density of $(S,T)$ given $A = a$ has the expression
\begin{equation}
\kappa(s, t|a) = c_1(a) s^{n - 1}\prod_{i = 1}^{n} f_0(s(a_i + x_i^T t)), \mbox{\ \ \ \ } s>0 \mbox{\ and \ } t \in \Bbb R^p,
\label{eq:kst}
\end{equation}
where $c_1(a)$ is a normalizing constant depending on $a$. Denote
by $g_{T|A}(\cdot|a)$ the conditional density of $T$ given $A=a$.
Then, for $t\in{\Bbb R}^p$ and ${\cal T}\subset{\Bbb R}^p$, the
integrals $g_{T|A}(t|a) = \int_0^{\infty}\!\!\!\kappa(s, t|a)
\,ds$ and $G_{T|A}({\cal T}|a) = \int_{t\in{\cal T}}
g_{T|A}(t|a)\, dt$ can be approximated by either Monte Carlo or
numerical integration if $f_0$ is known, with increasing
computational cost as $p$ increases. When $f_0$ is unspecified, we
note I-sufficiency of $A$ for $f_0$ and propose estimating $f_0$
by a kernel density estimate based on $a$: $\hat{f}_{h}(z|a)=(n
h)^{-1}\sum_{i=1}^{n}k\left((z-a_i)/h\right)$, where $k$ is a
kernel function and $h>0$ is the bandwidth. This leads to
nonparametric estimates $\hat{G}_{T|A}$ and $\hat{g}_{T|A}$ of
$G_{T|A}$ and $g_{T|A}$ respectively, which can again be
approximated by either Monte Carlo or numerical integration
methods. We term this the ``plug-in'' (PI) approach to distinguish
it from the ``residual bootstrap'' (RB) approach introduced
earlier to unconditional inference. The use of studentized
residuals $a$ in its derivation guarantees that $\hat f_h(z|a)$
has unit scale asymptotically. Under symmetry of $f_0$, it might
be beneficial in practice to use in place of $\hat{f}_h$ its
symmetrized version, $\tilde{f}_h(z|a)=(\hat f_h(z|a) + \hat
f_h(-z|a))/2$.

Under the regression model, the distribution and density of $U$ conditional on $\tilde A=\tilde a$ are given,
for ${\cal U}\subset{\Bbb R}^p$ and $u\in{\Bbb R}^p$, by
$G_{U|\tilde A}({\cal U}|\tilde a) = \int_{u\in{\cal U}} g_{U|\tilde A}(u|\tilde a)\, du$ and
$g_{U|\tilde A}(u|\tilde a) = c_3(\tilde a) \prod_{i = 1}^{n} f(\tilde a_i + x_i\tp u)$
respectively, for some constant $c_3(\tilde a)$. If $f$ is unspecified, the PI
approach substitutes $f$ by $\hat{f}_h(\cdot|\tilde a)$ or
$\tilde{f}_h(\cdot|\tilde a)$ to yield plug-in estimates
$\hat G_{U|\tilde A}$ and $\hat g_{U|\tilde A}$, on which
conditional inference can be based.

\section{Theory}
We consider first the asymptotic behaviour of $G_{T|A}$ and
$G_{U|\tilde A}$, and then assess the PI approach by substituting
kernel estimates for $f$ and $f_0$. Take $\ell_0\equiv \log f_0$
and $\ell\equiv\log f$ and assume the following regularity
conditions.
\begin{enumerate}
\item[(D1)] $f_0$ is symmetric about 0 and positive on $[-C,C]$
for some $C>0$. \item[(D2)] $f_0$ has uniformly bounded continuous
derivatives up to order 3, with $f_0'''$ being Lipschitz
continuous. \item[(D3)] $\expn\,\varepsilon^2$,
$\expn\,\varepsilon^2\ell_0'(\varepsilon)^2$,
$\expn\,\varepsilon^2\ell_0''(\varepsilon)^2$ and
$\expn\,|\varepsilon^3\ell_0'''(\varepsilon)|$ are finite for
$\varepsilon\sim f_0$.
\end{enumerate}
We assume that $X=X_n=[x_{n,1},\ldots,x_{n,n}]\tp$ depends on $n$
and satisfies the following.
\begin{enumerate}
\item[(C1)] $X_n\tp X_n$ is positive definite for all $n$, and
$\Sigma\equiv\lim_{n\rightarrow\infty}n^{-1}X_n\tp X_n$ exists and
is positive definite. \item[(C2)] (Generalized Noether condition)
$\displaystyle\lim_{n\rightarrow\infty}\max_{1\leq i\leq
n}\left\{{ x_{n,i}}\tp(X_n\tp X_n)^{-1}x_{n,i}\right\}=0$.
\item[(C3)] $\sup_n\left\{n^{-1}\sum_{i=1}^n({x_{n,i}}\tp
x_{n,i})^{1+\eta}\right\}<\infty$ for some $\eta>0$.
\end{enumerate}
Note that (C1) and (C2) imply asymptotic normality of least
squares estimators of $\beta$: see Sen and Singer ((1993),
Section~7.2). The location model provides a trivial example that
satisfies (C1)--(C3). The following theorem derives the asymptotic
conditional distributions of $n^{1/2}T$ and $n^{1/2}U$.
\begin{thm}
\label{thm:thm1} Assume (C1)--(C3), (D1)--(D3) and that
$\hat\beta=\beta+O_p(n^{-1/2})$ and
$\hat\sigma=\sigma+O_p(n^{-1/2})$. Then
\begin{itemize}
\item[(i)] under the regression-scale model, ${\cal I}^{1/2}(n^{1/2}T-{\cal I}^{-1}\theta)$ is standard
normal conditional on $A$, up to order $O_p(n^{-1/2})$,
where ${\cal I}=  n^{-2} X_n\tp X_n \sum_{i = 1}^{n}\ell_0'(A_i)^2$ and $\theta=  n^{-1/2} \sum_{i = 1}^{n}x_{n,i}\ell_0'(A_i)$;
\item[(ii)] under the regression model, $\tilde{\cal
I}^{1/2}(n^{1/2}U-\tilde{\cal I}^{-1}\tilde\theta)$ is standard normal conditional on $\tilde A$,
up to order $O_p(n^{-1/2})$, where
$\tilde{\cal I}=  n^{-2} X_n\tp X_n \sum_{i = 1}^{n}\ell'(\tilde A_i)^2$ and $\tilde\theta=  n^{-1/2} \sum_{i = 1}^{n}x_{n,i}\ell'(\tilde A_i)$.
\end{itemize}
\end{thm}
We see from Theorem $\ref{thm:thm1}$ that the conditional
distributions of $n^{1/2}T$ and $n^{1/2}U$ admit normal
approximations with conditional means and covariance matrices
depending on the score functions $\ell'_0$ and $\ell'$. The proof
of Theorem~\ref{thm:thm1} suggests that the conditional covariance
matrices ${\cal I}^{-1}$ and $\tilde{\cal I}^{-1}$ equal, up to
order $O_p(n^{-1/2})$, the deterministic matrices $I^{-1}$ and
$\tilde{I}^{-1}$, where $I=n^{-1} X_n\tp X_n \int(\ell'_0)^2f_0$
and $\tilde{I}=n^{-1} X_n\tp X_n \int(\ell')^2f$, whereas the
conditional means ${\cal I}^{-1}\theta$ and $\tilde{\cal
I}^{-1}\tilde\theta$ have asymptotic unconditional distributions
$N(0,I^{-1})$ and $N(0,\tilde{I}^{-1})$, respectively. It follows
that exact unconditional inference about $\beta$ may not be
correct, not even to first order asymptotically, conditional on
the ancillary residuals. For example, an unconditionally exact
level $1-\alpha$ confidence set derived from $G_T$ has conditional
coverage converging in probability to the random limit
$\Phi_{I^{-1}}(\Theta_{1-\alpha}-Z)$ for $Z\sim N(0,I^{-1})$,
where $\Phi_\Lambda$ denotes the $p$-variate $N(0,\Lambda)$
distribution and $\Phi_{\cal K}(\Theta_{1-\alpha})=1-\alpha$ for
some covariance matrix ${\cal K}$. The only exception is when
$\hat\beta$ is the exact maximum likelihood estimator of $\beta$.

To validate the PI approach asymptotically, we assume that the
kernel function $k$ satisfies the following.
\begin{enumerate}
\item[(K1)] $k$ has support $[-c,c]$, for some $c>0$, and is
symmetric about 0. \item[(K2)] $k$ is twice differentiable with
$k''$ being Lipschitz continuous. \item[(K3)] there exists some
$q\geq 2$ such that $\int k=1$, $\int u^jk(u)\,du=0$ for
$j=1,\ldots,q-1$, and $\int u^qk(u)\,du\neq 0$.
\end{enumerate}
First-order approximation of the PI approach amounts to
substitution of $\hat{f}_h(\cdot|A)$ or $\tilde{f}_h(\cdot|A)$ for
$f_0$ in the score $\ell'_0$ that defines the conditional normal
mean ${\cal I}^{-1}\theta$ and covariance matrix ${\cal I}^{-1}$
of $n^{1/2}T$. We consider a slightly different score estimator in
the theoretical development below. This simplifies the proof and
is asymptotically equivalent to the original PI proposal. Denote
by $A_{-i}$ the ancillary statistic $A$ with $A_i$ excluded, for
$i=1,\ldots,n$. Define, for $i=1,\ldots,n$ and $m=0,1,\ldots\:$,
the ``leave-one-out'' kernel estimator of $f_0^{(m)}$ by $\hat
f^{(m)}_{h}(z|A_{-i}) = ((n-1)h^{m+1})^{-1}\sum_{j \not= i}
k^{(m)}((z - A_j)/h)$. Symmetry of $f_0$ motivates an
anti-symmetrized leave-one-out estimate of $\ell_0'(A_i)$ given by
\[
\hat\ell'_{h_0, h_1}(A_i|A_{-i}) =2^{-1}\left\{\hat f'_{h_1}(A_i|A_{-i})/\hat f_{h_0}(A_i|A_{-i}) -\hat f'_{h_1}(-A_i|A_{-i})/\hat f_{h_0}(-A_i|A_{-i})\right\},
\]
for bandwidths $h_0,h_1>0$. This leads to estimators of $\theta$
and ${\cal I}$, given by $\theta^\dagger= n^{-1/2} \sum_{i =
1}^{n}x_{n,i}\hat\ell'_{h_0, h_1}(A_i|A_{-i})$ and ${\cal
I}^\dagger=n^{-2} X_n\tp X_n \sum_{i = 1}^{n}\hat\ell'_{h_0,
h_1}(A_i|A_{-i})^2$, respectively. Similar steps lead to estimates
$\tilde{\theta}^\dagger$ and $\tilde{\cal I}^\dagger$ of
$\tilde\theta$ and $\tilde{\cal I}$, respectively, under the
regression model. The following theorem concerns consistency of
the above estimators.
\begin{thm}
\label{thm:thm2} Assume (K1)--(K3), the conditions in
Theorem~\ref{thm:thm1} and that $h_m\rightarrow 0$ and
$nh_m^{2m+3}\rightarrow\infty$, $m=0,1$. Then
\begin{itemize}
\item[(i)] under the regression-scale model, ${\cal I}^\dagger={\cal I}+O_p(\delta_1)={\cal I}+o_p(1)$ and
$\theta^\dagger=\theta+O_p(\delta_2)=\theta+o_p(1)$;
\item[(ii)] under the regression model, $\tilde{\cal I}^\dagger=\tilde{\cal I}+O_p(\delta_1)=\tilde{\cal I}+o_p(1)$ and
$\tilde{\theta}^\dagger=\tilde{\theta}+O_p(\delta_2)=\tilde{\theta}+o_p(1)$,
\end{itemize}
where $\delta_1=h_0^q+h_1^q+n^{-1/2}(h_0^{-1/2}+h_1^{-3/2})$ and $\delta_2=h_0^q+h_1^q+n^{-1/2}(h_0^{-3/2}+h_1^{-5/2})$.
\end{thm}
Theorems~\ref{thm:thm1} and \ref{thm:thm2} together justify the PI
approach asymptotically and derive the valid orders of the
bandwidths involved. Note that the conditional distributions of
$n^{1/2}T$ and $n^{1/2}U$ can be estimated consistently by
$N({\cal I}^{\dagger-1}\theta^\dagger,{\cal I}^{\dagger-1})$ and
$N(\tilde{\cal I}^{\dagger-1}\tilde\theta^\dagger,\tilde{\cal
I}^{\dagger-1})$, respectively, provided that $h_0,h_1\rightarrow
0$ and $nh_0^{3},nh_1^{5}\rightarrow \infty$. We term this the
``normal approximate plug-in'' (NPI) approach to distinguish it
from the PI approach which directly simulates from, or numerically
evaluates, $\hat{G}_{T|A}$ and $\hat{G}_{U|\tilde A}$. The normal
approximation error can be kept to a minimum of order
$O_p(n^{-q/(5+2q)})$ by setting $h_1\propto n^{-1/(5+2q)}$,
$h_0=O(n^{-1/(5+2q)})$ and $h_0^{-1}=O(n^{5/(15+6q)})$. Park
(1993) introduced trimming constants to the estimated score $\hat
\ell_{h_0,h_1}'$ to correct for its occasional erratic behaviour.

\noindent {\bf Remark.} Adaptive estimation constructs
asymptotically efficient estimators by substituting nonparametric
score estimates in a one-step maximum likelihood approximation.
Stone (1975) considered adaptive estimation under the symmetric
location model. Bickel (1982) extended the construction to linear
models. Under our  regression setup,  the adaptive estimator built
upon an equivariant regression estimator has conditional and
unconditional distributions equivalent to first order, and can be
viewed as an equivariant regression estimator with conditional
mean recentered at the true regression parameter. This connection
implies asymptotic equivalence of adaptive estimation and the NPI
approach, suggesting that the latter can be approximated by
unconditional inference based on adaptive estimators. Many
nonparametric methods, such as the bootstrap, that are intended
mainly for unconditional inference are readily available for
estimation of such unconditional distributions.

\section{Robustness and configural polysampling}
Morgenthaler and Tukey (1991) suggest a global, finite-sample,
notion of robustness that pays due attention to ancillarity. Their
method, known as configural polysampling, makes robust inference
by conditioning on an ancillary configuration of the observed data
under a confrontation of rival parametric models. Its dissociation
from asymptotic reasoning makes the method attractive for finite
samples and distinct from such conventional devices as the
influence function and the breakdown point. Morgenthaler (1993)
specializes it to linear models and develops computationally
simple procedures for robust estimation.

A key ingredient to configural polysampling is the choice of a
confrontation pair $(\cal F,G)$, where $\cal F$ and $\cal G$
denote extremes, in a spectrum of error distributions of practical
interest, under which inference is done separately and the
resulting analyses combined in an optimal way. The approach can be
generalized to deal with more than two distributions in the
confrontation. Morgenthaler and Tukey (1991) suggest taking $\cal
F$ and $\cal G$ to be the normal and slash distributions to
encompass a spectrum ranging from light- to heavy-tailed
distributions. To fix ideas, consider estimation of  $\beta$ by an
equivariant estimator $V$, such that $V(Y)=\hat\beta+\hat\sigma
V(A)$. When $f_0={\cal F}$, the conditional mean squared error
(cMSE) of $V$ given $A$ is minimized at $V(A)=V_{\cal
F}(A)\equiv-\Bbb E_{\cal F}[S^2 T|A]/\Bbb E_{\cal F}[S^2|A]$,
leading to Pitman's (1939) famous estimator, an early example of
optimal estimation driven by the conditionality principle. Thus we
may write, for an arbitrary equivariant estimator $V$,
$\mbox{cMSE}_{\cal F}(V|A)=\mbox{cMSE}_{\cal F}(V_{\cal
F}|A)+\sigma^2 \Bbb E_{\cal F}[S^2|A](V(A)-V_{\cal F}(A))^2$.
Morgenthaler and Tukey (1991) select a ``bioptimal'' $V$ by
minimizing $P_{\cal F}\times \mbox{cMSE}_{\cal F}(V|A)+P_{\cal
G}\times\mbox{cMSE}_{\cal G}(V|A)$, for a pair of shadow prices
$P_{\cal F}$ and $P_{\cal G}$. Alternatively, a minimax estimator
$V$ of $\beta$ can be obtained  by minimizing the maximum of
$\mbox{cMSE}_{\cal F}(V|A)$ and $\mbox{cMSE}_{\cal G}(V|A)$, which
often amounts to solving the equation $\mbox{cMSE}_{\cal
F}(V|A)=\mbox{cMSE}_{\cal G}(V|A)$. The regression model can be
treated similarly. In confidence interval problems one may, for
example, minimize the conditional mean interval length subject to
correct unconditional coverages under $\cal F$ and $\cal G$. In
general, configural polysampling  fine-tunes statistical
procedures  to achieve simultaneous efficiency over a spectrum of
distributions determined by $(\cal F, \cal G)$. It can be
generalized with data-driven choices of $(\cal F,\cal G)$, thereby
robustifying the inference procedure in a global sense. We
envisage confrontations $(\cal F,\cal G)$ which reflect practical
concerns in robust statistical inference. For example,  we may
confront parametric with nonparametric approaches, unconditional
with conditional approaches, asymptotic approximation with
finite-sample methods, small with large bandwidths in any
kernel-based approach, or any two competing nonparametric
approaches. In these possible confrontations, our PI or NPI
approaches can play a prominent role in robustifying the inference
outcome specific to the observed ancillary configuration. Further
empirical evidence is presented in Section~5.2.

\section{Empirical studies}

\subsection{Confidence intervals}
Our first study compared the conditional coverage probabilities of
the PI and NPI intervals with those of the exact unconditional and
RB intervals. We considered the location model with $p=1$ and
$\beta=1$, and took $f$ to be the Student's $t_5$ density, which
satisfies (D1)--(D3). The ``conditional" samples, all subject to a
common observed value of $\tilde A$, were obtained by rejection
sampling. Three sample sizes, $n=15$, 30 and 100, were considered.
The nominal level $1-\alpha$ was chosen to be
$0.90,0.91,\ldots,0.99$. Each conditional coverage was estimated
from 5000 ``conditional" samples. Construction of the PI and NPI
intervals was based on 5000 samples drawn from $\hat{G}_{U|\tilde
A}$ and its normal approximation, respectively. The RB interval
was based on 1000 bootstrap samples and the exact unconditional
interval on 5000 samples drawn from $f$ itself. The kernel
function $k$ was taken to be the standard normal density.

The objective of this study is to demonstrate the importance of
conditioning and the effectiveness of PI and NPI in constructing
conditional confidence intervals. Despite its importance in
practice, the issue of bandwidth selection is not our main
interest and we set $h=1$ throughout the study, the best choice in
a pilot study done on four different sets of ancillary residuals.
Conventional methods for practical bandwidth selection include the
normal referencing rule,  cross-validation and the (conditional)
bootstrap. Alternatively, an innovative approach can be based on
configural polysampling under a confrontation of two extreme
choices of bandwidth. This will be illustrated in Section~5.2. For
the NPI approach we used the true $f$, rather than its kernel
estimate, for computing $\ell'$ in order to examine the effects on
conditional coverages due exclusively to normal approximation.

Figure~\ref{fig:n15} plots the conditional coverage errors against $1-\alpha$
for $n=15$ for four different sets of $\tilde A$, chosen
specifically such that the exact unconditional
intervals undercover in two cases and overcover in the other two.
We see that the exact unconditional interval has very large
conditional coverage error compared to the two plug-in approaches,
except for the fourth case where it outperforms the NPI approach. Surprisingly, the RB interval yields more accurate coverage than does the exact unconditional interval,
although the former is designed primarily for estimating the latter.
It is evident that $U$ has very different unconditional and
conditional distributions given our choices of $\tilde A$. The PI
approach works effectively for all four choices of
$\tilde{A}$. Inferior in general to the PI intervals, NPI
nevertheless corrects the
exact unconditional interval to some extent, although the
correction is less remarkable when the unconditional
interval overcovers. Similar conclusions are observed for
$n=30$ and 100. We also investigated choices of $\tilde A$ given which the exact unconditional interval is conditionally accurate.
The results, not shown in this report, suggest that both the PI and NPI intervals remain, as expected, accurate in those cases.

\subsection{Robust conditional estimation}
The second study illustrates applications of the PI approach in
configural polysampling procedures for robust conditional
inference. We considered three types of confrontation pairs, all
reflecting genuine practical concerns: (i) the normal versus the
slash distributions; (ii) the least squares method versus the PI
approach based on bandwidth $h=Cn^{-1/9}$, a multiple of the
optimal order; and (iii) the PI approach based on contrasting
bandwidths $h_a$ and $h_b$. Note that (i) was conceived by
Morgenthaler and Tukey (1991) for achieving robustness across
symmetric, unimodal, distributions of different tail behaviour.
Case (ii) contrasts conditional with unconditional inferences.
Case (iii) suggests a practical robust solution, which respects
the conditionality principle,  to the problem of bandwidth
selection in the PI method. In the study we set
$C=0.1,0.5,1.0,1.5,2.0,2.5$ and $h_a=0.1,h_b=2.0$. The kernel $k$
was taken to be the standard normal density.

We considered again a location model and compared the mean squared
error of minimax location estimates obtained under different
confrontations. The least squares estimate, the sample mean, was
also included for comparison. Given a fixed set of residuals, we
generated 100,000 ``conditional'' random samples of sizes $n=15$
and $30$ from each of six different distributions: the Student's
$t_1$, the normal mixture $\frac{1}{2}N(-3,1)+\frac{1}{2}N(3,1)$
and the centered beta distributions with support $[-5,5]$ and
shape parameters $(1/2,1/2),(2,2),(1/2,2)$ and $(2,1/2)$, among
which the $t_1$ and $\beta(2,2)$ densities have bell-like shapes
and can be deemed to lie within the normal-slash spectrum. We are
here not so much concerned with asymptotic validity as interested
in robustness against model departures in a broad context. Indeed,
all six distributions except the normal mixture fail to satisfy
(D3).

Table~\ref{tab:poly} reports the cMSE's of
the various estimates, obtained by
averaging over the conditional samples generated from each
distribution. We see that confrontation types (ii) and (iii) give
remarkably small cMSE compared to (i), which is even less accurate than the
unconditional least squares estimate under distributions outside
the normal-slash spectrum. Confrontation type
(ii) outperforms (i) under all choices of $C$ and most of the
underlying distributions except $t_1$, under which use of
large $C$ in (ii) gives results comparable to (i).
Particularly encouraging are the results obtained using
confrontation (iii), which returns an accurate, robustified PI
estimate for which the bandwidth is implicitly selected from
candidate values lying between $h_a$ and $h_b$.

\subsection{A real data example}
DiCiccio (1988) and Sprott (1980, 1982) made conditional inference
about a real location parameter $\beta$ by fitting a
location-scale model with $t_\lambda$ error to Darwin's data
(Fisher (1960), P.37) on 15 height differences between cross- and
self-fertilized plants. We removed the $t_\lambda$ assumption, set
$\hat\beta$ to be (I) the sample mean and (II) the sample median,
both being location equivariant, and constructed $95\%$ two-sided
RB, PI and NPI intervals for $\beta$ in both cases. The RB
interval was based on 50,000 bootstrap samples. The NPI interval
was built on the anti-symmetrized leave-one-out score estimate,
for which the bandwidths $h_0$ and $h_1$ were fixed to be (give
the actual number, not formula) using the normal referencing rule.

For (I), we calculated the RB and NPI intervals to be $(2.46,
39.43)$ and $(8.78, 40.80)$, respectively, and the PI intervals to
be $(17.51, 24.45)$, $(11.33, 35.84)$, $(10.42, 39.12)$, $(8.38,
42.15)$, $(4.86, 44.44)$ and $(2.24, 45.51)$ based on bandwidths
$h=mh_0$, for $m=0.2,0.5,0.7,1.0,1.3,1.5$, respectively. The
results are in agreement with DiCiccio's (1988) and Sprott's
(1980, 1982) findings, suggesting plausibility of their Student's
$t$ error assumption. The case (II) gives similar results except
that the endpoints are shifted slightly to the right, in general.

\section{Conclusion}
We establish consistency of the PI approach to conditional
inference and derive sufficient bandwidth orders. The NPI approach
provides a computationally convenient normal approximation to it.
Effectiveness of the approaches is confirmed by empirical
findings. The computational cost of PI depends on the dimension
$p$ and the efficiency with which we can simulate from $\hat
g_{T|A}$ or $\hat{g}_{U|\tilde A}$. The computing times for both
plug-in approaches were found to be within seconds under the
location model considered in Section 5.1.

Incorporation of the plug-in approaches into confrontations extends configural polysampling to
the nonparametric realm, rendering the resulting conditional inference an extra dimension of robustness.
When applied to a confrontation of two extreme bandwidths, the technique suggests an innovative solution, which observes
the conditionality principle, to   bandwidth selection
in practical applications of the PI approaches.
We remark that confrontations of more than two specifications of error density can be considered in configural polysampling
to further robustify the inference outcome, although then the minimax algorithm is necessarily more computationally involved.

\section{Appendix}
\subsection{Proof of Theorem~\ref{thm:thm1}}
Under the regression-scale model, we deduce, by a Taylor expansion
of (\ref{eq:kst}) in powers of $n^{-1/2}$, that the density of
$n^{1/2}(\log S,T)$ conditional on $A$
 is proportional, up to $O_p(n^{-1/2})$, to the product of the $N({\cal J}^{-1}\psi,{\cal J}^{-1})$ and $N({\cal I}^{-1}\theta,{\cal I}^{-1})$ density
functions, where ${\cal J}= -\, n^{-1} \sum_{i = 1}^n\left\{A_i
\ell_0'(A_i)+A_i^2 \ell_0''(A_i)\right\}$ and $\psi=
n^{-1/2}\sum_{i = 1}^{n}\left( A_i \ell_0'(A_i) + 1\right)$. This
proves part (i). Part (ii) follows by similar, but simpler
arguments.

\subsection{Proof of Theorem~\ref{thm:thm2}}
Note that Linton and Xiao's (2001) Lemma~2 can be adapted to deduce that
\begin{equation}
\hat f_{h}^{(m)}(\pm A_i|A_{-i}) = f^{(m)}_0(\pm A_i) + O_p(h^q +
n^{-1/2} h^{-m - 1/2}),\;\;m=0,1, \label{eq:Xiao}
\end{equation}
uniformly in $i\in\{1,\ldots,n\}$. It follows that
$\hat\ell'_{h_0, h_1}(A_i|A_{-i})=\ell'_0(A_i)+O_p(\delta_1)$, and
hence the result for ${\cal I}^\dagger$.

Define $\delta^\pm_{im}=\pm\left\{\hat f_{h_m}^{(m)}(\pm\epsilon_i|A_{-i})-\hat f_{h_m}^{(m)}(\pm A_i|A_{-i})\right\}$.
That $\hat\beta$ and $\hat\sigma$ are $n^{1/2}$-consistent
implies that
$\mu\equiv\log(\hat\sigma/\sigma)$ and $\tau\equiv\hat\beta/\hat\sigma-\beta/\sigma$ are both $O_p(n^{-1/2})$.
Write $\bar{x}=\sum_{i=1}^nx_{n,i}/n$. Conditioning on $\epsilon_i$,
standard asymptotic theory yields
\begin{eqnarray}
\lefteqn{(n-1)^{-1}h_m^{-m-2}\sum_{j\neq
i}k^{(m+1)}((\pm\epsilon_i-\epsilon_j)/h_m)(\pm x_{n,i}-x_{n,j})}\nonumber\\
&=&  f^{(m+1)}_0(\pm\epsilon_i)(\pm
x_{n,i}-\bar{x})+O_p(h_m^q+n^{-1/2}h_m^{-m-3/2}),\nonumber
\end{eqnarray}
\begin{eqnarray}
\lefteqn{(n-1)^{-1}h_m^{-m-2}\sum_{j\neq
i}k^{(m+1)}((\pm\epsilon_i-\epsilon_j)/h_m)(\pm \epsilon_i-\epsilon_j)}\nonumber\\
&=& -\,(m+1)
f^{(m)}_0(\pm\epsilon_i)+O_p(h_m^q+n^{-1/2}h_m^{-m-1/2}),\nonumber
\end{eqnarray}
so that
\begin{eqnarray}
\delta^\pm_{im}&=&\pm f_0^{(m+1)}(\pm\epsilon_i)(\pm
x_{n,i}-\bar{x})\tp(\tau+\mu\beta/\sigma)\mp (m+1)\mu
f_0^{(m)}(\pm\epsilon_i)\nonumber\\
&&\mbox{\ \hspace{3cm}}+O_p(n^{-1/2}h_m^q+n^{-1}h_m^{-m-3/2}).
\label{eq:delta2}
\end{eqnarray}
Noting (\ref{eq:delta2}), and that (\ref{eq:Xiao}) also holds if
$\pm A_i$ is replaced by $\pm\epsilon_i$, we have
\begin{eqnarray}
\lefteqn{\hat\ell'_{h_0, h_1}(A_i|A_{-i})}\nonumber\\
& =& \hat f'_{h_1}(\epsilon_i|A_{-i})/\hat
f_{h_0}(\epsilon_i|A_{-i}) -
\left[f''_0(\epsilon_i)x_{n,i}\tp(\tau+\mu\beta/\sigma)-2\mu
f'_0(\epsilon_i)\right]/f_0(\epsilon_i)\nonumber\\
&& + \left[f'_0(\epsilon_i)x_{n,i}\tp(\tau+\mu\beta/\sigma)-\mu
f_0(\epsilon_i)\right]f'_0(\epsilon_i)/f_0(\epsilon_i)^2+O_p\left(n^{-1/2}\delta_2\right).
\label{eq:score}
\end{eqnarray}
Expanding $A_i$ about $\epsilon_i$, we have
\begin{eqnarray}
\ell'_0(A_i)&=&f'_0(\epsilon_i)/f_0(\epsilon_i) -
\left[x_{n,i}\tp(\tau+\mu\beta/\sigma)+\mu\epsilon_i
\right]f''_0(\epsilon_i)/f_0(\epsilon_i)\nonumber\\
&& + \left[x_{n,i}\tp(\tau+\mu\beta/\sigma)+\mu\epsilon_i
\right]f'_0(\epsilon_i)^2/f_0(\epsilon_i)^2+O_p(n^{-1}).
\label{eq:true.score}
\end{eqnarray}
Symmetry of $f_0$ and (D3) together imply that
$n^{-1/2}\sum_{i=1}^nx_{n,i}f'_0(\epsilon_i)/f_0(\epsilon_i)$,
$n^{-1/2}\sum_{i=1}^nx_{n,i}\epsilon_i
f''_0(\epsilon_i)/f_0(\epsilon_i)$, and
$n^{-1/2}\sum_{i=1}^nx_{n,i}\epsilon_if'_0(\epsilon_i)^2/f_0(\epsilon_i)^2$
are all of order $O_p(1)$. It then follows from (\ref{eq:score})
and (\ref{eq:true.score}) that
\begin{equation}
\theta^\dagger-\theta=n^{-1/2}\sum_{i=1}^nx_{n,i}\left\{\hat
f'_{h_1}(\epsilon_i|A_{-i})/\hat f_{h_0}(\epsilon_i|A_{-i}) -
f'_0(\epsilon_i)/f_0(\epsilon_i)\right\}+O_p(\delta_2).
\label{eq:theta.diff}
\end{equation}
The proof of Linton and Xiao's (2001) Theorem~1 can be adapted to
show that the first term in (\ref{eq:theta.diff}) has
order $O_p(\delta_1)$,
which can be absorbed into
$O_p(\delta_2)$. This
completes the proof of (i). Part (ii) follows by similar arguments.

\pagebreak
\pagestyle{empty}
\begin{center}
\input epsf\vspace{5mm}
\begin{figure}[h]
\caption{Conditional coverage errors of exact unconditional, PI,
NPI and RB intervals, for $n=15$.}
\centering \leavevmode \epsfxsize=13cm \epsfbox{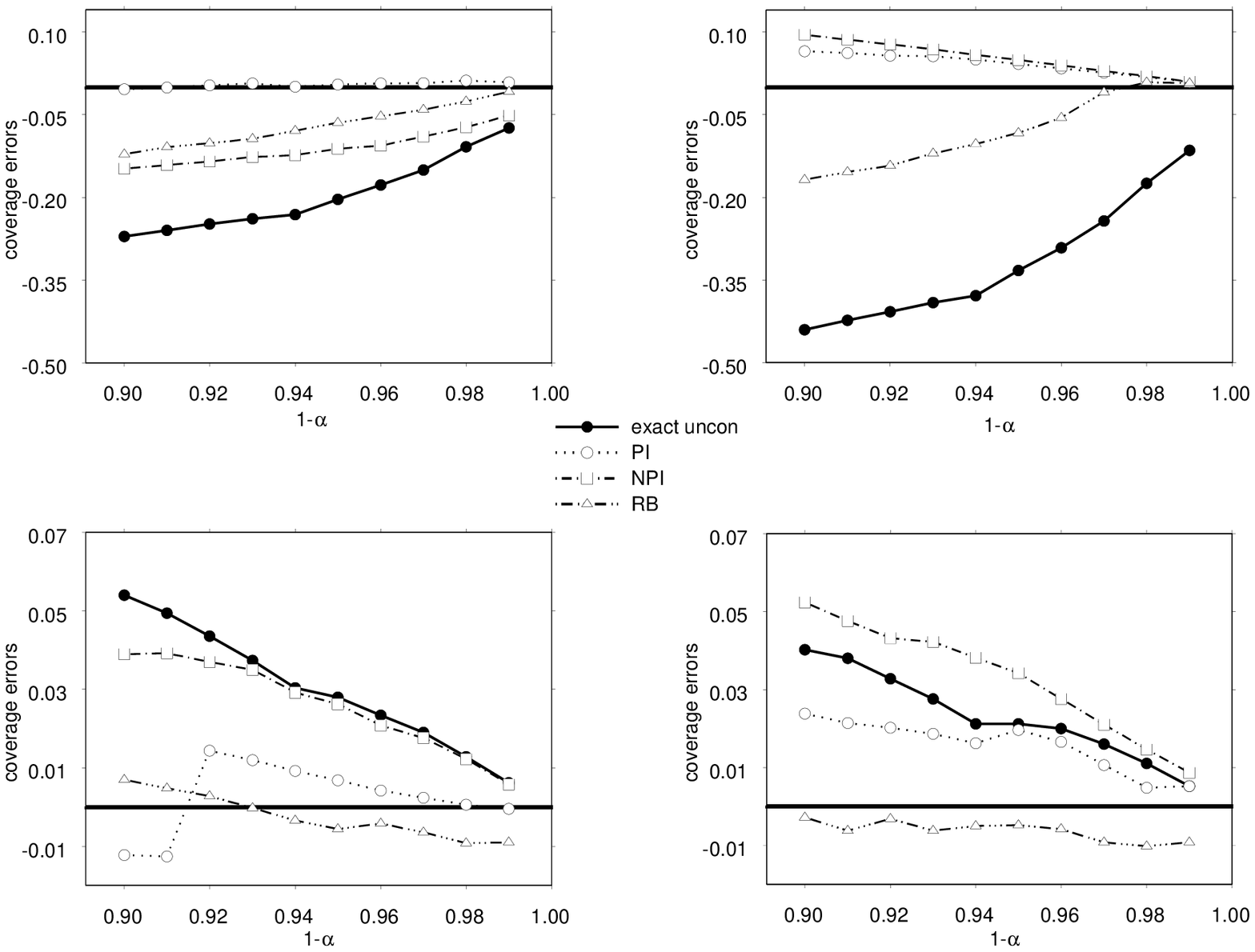}\\
\label{fig:n15}
\end{figure}
\end{center}
\oddsidemargin 0.6cm \evensidemargin 0.6cm
\renewcommand{\baselinestretch}{1.2}
\begin{table}[t]
\caption{Conditional mean squared errors of least squares estimates (LS) and minimax estimates obtained under confrontations (i) normal vs slash, (ii)
LS vs PI, (iii) $h_a=0.1$ vs $h_b=2.0$ in PI\@.}
\label{tab:poly}
\begin{center}\
\small
\begin{tabular}{|ll|cccccc|}
\hline
&&\multicolumn{6}{c|}{Error distribution} \\ \cline{3-8}
&& centred &  centred & centred & centred && $\frac{1}{2}N(-3,1)$ \\
\multicolumn{2}{|l|}{Confrontation} & $\beta(1/2,1/2)$ &$\beta(2,2)$ &$\beta(1/2,2)$ & $\beta(2,1/2)$ & $t_1$ & $+\frac{1}{2}N(3,1)$\\ \hline
\multicolumn{2}{|l|}{\boldmath{$n=15$}}&&&&&&\\
(i) & normal vs slash & 4.3403 & 0.6814 & 1.5399 & 1.5743 & 0.5745 &
2.7075\\
(ii) & LS vs PI ($C=0.1$)&  1.3105 & 0.4174 & 0.7947 & 0.7979 & 2.0863 & 1.0709\\
(ii) & LS vs PI ($C=0.5$)&  1.2326 & 0.4174 & 0.9643 & 0.9674 & 2.0863 & 0.4513\\
(ii) & LS vs PI ($C=1.0$)&  0.4206 & 0.3869 & 1.3331 & 1.3352 & 0.9975 & 0.4351\\
(ii) & LS vs PI ($C=1.5$)&  0.2234 & 0.3790 & 1.4512 & 1.4540 & 0.7163 & 0.3631\\
(ii) & LS vs PI ($C=2.0$)&  0.1995 & 0.3889 & 1.5092 & 1.5118 & 0.6033 & 0.3921\\
(ii) & LS vs PI ($C=2.5$)&  0.2952 & 0.4017 & 1.1576 & 1.1633 & 0.5335 & 0.5787\\
(iii) & $h_a=0.1$ vs $h_b=2.0$ &  0.3508 & 0.4057 & 1.0683 & 1.0670 & 0.4664
& 0.6532\\
& LS & 1.3126 & 0.4178 & 0.7963 & 0.7963 & 2.0837 &
1.0691\\ \hline
\multicolumn{2}{|l|}{\boldmath{$n=30$}}&&&&&&\\
(i) & normal vs slash & 1.6922 & 0.3884 & 0.3200 & 0.3216 & 0.5380 &
4.4011\\
(ii) & LS vs PI ($C=0.1$)&  0.2277 & 0.1946 & 0.0386 & 0.0391 & 1.9589 & 1.9654\\
(ii) & LS vs PI ($C=0.5$)&  0.1068 & 0.1796 & 0.0386 & 0.0391 & 1.9589 & 1.9654\\
(ii) & LS vs PI ($C=1.0$) & 0.0116 & 0.1828 & 0.0171 & 0.0169 & 1.2311 & 1.2539\\
(ii) & LS vs PI ($C=1.5$)&  0.0049 & 0.1801 & 0.0125 & 0.0120 & 0.8892 & 1.0360\\
(ii) & LS vs PI ($C=2.0$) & 0.0010 & 0.1801 & 0.0187 & 0.0189 & 0.7061 & 0.9042\\
(ii) & LS vs PI ($C=2.5$) &  0.0071 & 0.1834 & 0.0276 & 0.0279 & 0.6188 & 0.9591\\
(iii) & $h_a=0.1$ vs $h_b=2.0$ & 0.0186 & 0.1867 & 0.0321 & 0.0328 & 0.5566
& 1.1533\\
& LS & 0.2283 & 0.1947 & 0.0389 & 0.0389 & 1.9571 &
1.9636\\
\hline
\end{tabular}
\end{center}
\end{table}


\begin{thebibliography}{99}

\bibitem{}
Barnard, G.A. (1981). The conditional approach to robustness. {\em
Statistics and Related Topics\/}, 235--241. North-Holland, New
York.

\bibitem{}
Barnard, G.A. (1983). Pivotal inference and the conditional view
of robustness. {\em Scientific Inference, Data Analysis, and
Robustness\/}, 1--8. Academic Press, New York.

\bibitem{}
Barndorff-Nielsen, O. (1978). {\em Information and Exponential
Families in Statistical Theory\/}. John Wiley, New York.

\bibitem{}
Bickel, P.J. (1982). On adaptive estimation. {\em Ann.\
Statist.\/} {\bf 10}, 647--671.

\bibitem{}
Cox, D.R. and Hinkley, D.V. (1974). {\em Theoretical
Statistics\/}. Chapman and Hall, London.

\bibitem{}
 DiCiccio, T.J. (1988). Likelihood inference for linear
regression models. {\em Biometrika\/} {\bf 75}, 29--34.

\bibitem{}
 DiCiccio, T.J., Field, C.A. and Fraser, D.A.S. (1990).
Approximations of marginal tail probabilities and inference for
scalar parameters. {\em Biometrika\/} {\bf 77}, 77--95.

\bibitem{}
 Fisher, R.A. (1934). Two new properties of mathematical likelihood.  {\em Proc.\ Roy.\ Soc.\ A\/} {\bf 144}, 285--307.

\bibitem{}
 Fisher, R.A. (1935). The logic of inductive inference. {\em J.\ Roy.\ Statist.\ Soc.\/}  {\bf 98}, 39--54.

\bibitem{}
 Fisher, R.A. (1960). {\em The Design of Experiments.\/}
Oliver and Boyd, Edinburgh.

\bibitem{}
 Fraser, D.A.S. (1976). Necessary analysis and adaptive inference. {\em J.\ Amer.\ Statist.\ Assoc.\/}  {\bf 71}, 99--110.

\bibitem{}
 Fraser, D.A.S. (1979). {\em Inference and Linear Models.\/}
McGraw-Hill, New York.

\bibitem{}
 Hinkley, D.V. (1978). Likelihood inference about location
and scale parameters. {\em Biometrika\/}  {\bf 65}, 253--261.

\bibitem{}
 Hinkley, D.V. (1983). Can frequentist inferences be very
wrong? A conditional ``Yes''. {\em Scientific Inference, Data
Analysis, and Robustness\/},  1--8. Academic Press, New York.

\bibitem{}
 J\o rgensen, B. (1993). A review of conditional inference: is there a universal definition of nonformation? {\em Bull.\ Int.\ Statist.\ Inst.\/} {\bf 55}, 323--340.

\bibitem{}
 Lehmann, E.L. (1981). An interpretation of completeness and
Basu's Theorem. {\em J.\ Amer.\ Statist.\ Assoc.\/}  {\bf 76},
335--339.

\bibitem{}
 Linton, O. and Xiao, Z. (2001). Second-order approximation for adaptive regression estimators. {\em Econometric Theory\/} {\bf 17}, 984--1024.

\bibitem{}
 Morgenthaler, S. (1993). Robust tests for linear models. {\em Statistical Sciences and Data Analysis\/} (Tokyo, 1991), 97--107.
Utrecht, VSP.

\bibitem{}
 Morgenthaler, S. and Tukey, J.W. (1991). {\em Configural
Polysampling: A Route to Practical Robustness\/}. John Wiley, New
York.

\bibitem{}
 Park, B.U. (1993). A cross-validatory choice of smoothing parameter in adaptive location estimation. {\em J.\ Amer.\ Statist.\ Assoc.\/} {\bf 88}, 848--854.

\bibitem{}
 Pitman, E.J.G. (1939). The estimation of location and scale parameters
of a continuous population of any given form. {\em Biometrika\/} {\bf 30}, 391--421.

\bibitem{}
 Seifu, Y., Severini, T.A. and Tanner, M.A. (1999) Semiparametric Bayesian inference for regression models. {\em Can.\ J. Statist.\/} {\bf 27}, 719--734.

\bibitem{}
 Sen, P.K. and Singer, J.M. (1993). {\em Large Sample Methods
in Statistics: an Introduction with Applications\/}. Chapman and
Hall, New York.

\bibitem{}
 Severini, T.A. (1992). Conditional robustness in location
estimation. {\em Biometrika\/} {\bf 79}, 69--79.

\bibitem{}
 Severini, T.A. (1994). Nonparametric conditional inference for a location parameter. {\em J.\ Roy.\ Statist.\ Soc.\ B\/}  {\bf 56}, 353--362.

\bibitem{}
 Severini, T.A. (1996). Measures of the sensitivity of
regression estimates to the choice of estimator. {\em J.\ Amer.\
Statist.\ Assoc.\/} {\bf 91}, 1651--1658.

\bibitem{}
 Sprott, D.A. (1980). Maximum likelihood in small samples:
Estimation in the presence of nuisance parameters. {\em Biometrika
\/} {\bf 67}, 515--523.

\bibitem{}
 Sprott, D.A. (1982). Robustness and maximum likelihood
estimation. {\em Comm.\ Statist.\ A} {\bf 11}, 2513--2529.

\bibitem{}
 Stone, C. (1975). Adaptive maximum likelihood estimation of a location parameter. {\em Ann.\ Statist.\/} {\bf 3}, 267--284.

\end{thebibliography}
\end{document}